\documentclass{elsarticle}
\usepackage{xcolor}
\usepackage{bm}
\usepackage{graphicx}
\usepackage{amsfonts}
\usepackage{amsmath}

\usepackage{lineno,hyperref}

\begin{document}

\begin{frontmatter}

\title{On the statistical properties of viral misinformation in online social media}

\author{Alessandro Bessi}
\corref{mycorrespondingauthor}
\cortext[mycorrespondingauthor]{Corresponding author}
\ead{bessi@isi.edu}

\address[]{University of Southern California, Information Sciences Institute, Marina del Rey, Los Angeles, CA, USA}
\address[mymainaddress]{IUSS Institute for Advanced Study, Pavia, ITALY}
\address[mysecondaryaddress]{IMT Institute for Advanced Studies, Lucca, ITALY}

\begin{abstract}
The massive diffusion of online social media allows for the rapid and uncontrolled spreading of conspiracy theories, hoaxes, unsubstantiated claims, and false news. Such an impressive amount of misinformation can influence policy preferences and encourage behaviors strongly divergent from recommended practices. In this paper, we study the statistical properties of viral misinformation in online social media. By means of methods belonging to Extreme Value Theory, we show that the number of extremely viral posts over time follows a homogeneous Poisson process, and that the interarrival times between such posts are independent and identically distributed, following an exponential distribution. Moreover, we characterize the uncertainty around the rate parameter of the Poisson process through Bayesian methods. Finally, we are able to derive the predictive posterior probability distribution of the number of posts exceeding a certain threshold of shares over a finite interval of time.
\end{abstract}

\begin{keyword}
misinformation \sep online social media \sep extreme value theory
\end{keyword}

\end{frontmatter}


\section{Introduction}
The wide availability of user-provided contents in online social media encourages the aggregation of people around common interests and narratives. The direct path from producers to consumers of contents drives the emergence of a disintermediated enviroment that is changing the way people become informed, interpret facts, and form their opinions \cite{brown2007word, kahn2004new, quattrociocchi2011opinions, quattrociocchi2014opinion, kumar2010dynamics}.

Unfortunately, such a disintermediation can facilitate the spreading of rumors, hoaxes, fake news, and conspiracy theories, that often arouse naive and awkward social responses on different topics, such as health, environment, national security, and politics \cite{RumorEbola1, RumorEbola2, RumorEbola3, RumorJadeHelm, RumorVaccine, RumorTrump}. In particular, conspiracy theories  simplify causation, reduce the complexity of reality, and are formulated in a way that is able to contain a certain level of uncertainty \cite{sunstein2009conspiracy, byford2011conspiracy, fine2005rumor, hogg2011extremism}. However, such an impressive amount of misinformation can influence policy preferences and encourage behaviors strongly divergent from recommended practices.

Since the World Economic Forum listed massive digital misinformation as one of the main threats to our society \cite{howell2013digital}, community-driven \cite{facebook2015} and algorithmic-driven \cite{qazvinian2011rumor, ciampaglia2015computational, resnick2014rumorlens, gupta2014tweetcred, almansour2014model, ratkiewicz2011detecting, dong2015knowledge} solutions have been proposed to counteract the pervasiveness of online misinformation. However, a part of the scientific community is skeptical about the real effectiveness of such solutions. Indeed, the community-driven approach proposed by Facebook -- where users can flag false contents to correct the newsfeed algorithm -- is controversial, because it raises fears that the free circulation of ideas may be threatened. Moreover, algorithmic-driven approaches may not be effective, since the acceptance of a claim (either substantiated or not) is heavily influenced by social norms and individual cognitive factors \cite{mocanu2015collective, bessi2015science,nyhan2014effective,javarone2014social,javarone2013perception,javarone2014network}. Indeed, recent works point out both the inefficacy of correcting false beliefs and the concrete risk of a backfire effect \cite{bessi2014social, zollo2015debunking, garrett2013undermining} from the usual and most committed consumers of conspiracy theories. In fact, false beliefs, once adopted by an individual, are rarely corrected \cite{garrett2013promise, meade2002explorations, koriat2000toward, ayers1998theoretical}.

More specifically, both the formation and the revision of beliefs are strongly affected by the communities wherein ideas and facts are debated \cite{zhu2010individual, frenda2011current}. Such a phenomenon is emphasized in online social networks, where users process information through a shared system of meaning \cite{bessi2015trend, zollo2015emotional} inside their echo chambers \cite{del2016spreading, bessi2015viral, bessi2016users,bessi2016personality}, making sense of facts in ways that are often biased toward self-confirmation. Indeed, recent studies show that increasing the exposure of users to unsubstantiated rumors increases their tendency to be credulous \cite{bessi2015science, bessi2014economy}, and that the content-selective exposure is the primary driver of content diffusion, and generates the formation of echo chambers \cite{del2016spreading}.

In this work, we study the statistical properties of viral misinformation in online social media. In particular, we apply methods of Extreme Value Theory -- a branch of statistics dealing with extreme deviations from the median of probability distributions -- to analyze a large dataset of posts published by Facebook pages supporting conspiracy theories and myth narratives. By means of an in-depth statistical analysis of the shares distribution and the application of the Peaks Over Threshold (POT) approach, we show that the number of extremely viral posts (e.g. $>250K$ shares) over time follows a homogeneous Poisson process, and that the interarrival times between such posts are independent and identically distributed, following an exponential distribution. Further, we characterize the uncertainty around the rate parameter of the Poisson process through Bayesian methods. Finally, we are able to derive the predictive posterior probability distribution of the number of posts exceeding a certain threshold of shares over a finite interval of time.

The relevance of our results is not necessarily limited to the field of computational social science coping with misinformation \cite{shao2016hoaxy, grimes2016viability, zubiaga2015towards}. Indeed, despite the prediction of extremely viral posts and rare events remains an hard task \cite{salganik2006experimental, watts2011everything}, we believe that both our findings and the methodology used herein may be of interest to the broader field of computational social science dealing with forecasting and tracking of viral contents and events \cite{cheng2014can, friggeri2014rumor, staiano2013exploring, hoang2012virality, hong2011predicting, jenders2013analyzing, yang2010predicting, coscia2014average, weng2013virality, zhou2016early}.

\section{Methods}

\subsection{Ethics Statement}
The entire data collection process has been carried out exclusively through the Facebook Graph API, which is publicly available. We used only public available data. The pages from which we downloaded data are public Facebook entities.

\subsection{Data Collection}
We analyzed 328 US public Facebook pages diffusing conspiratorial beliefs, myth narratives, and controversial information, usually lacking supporting evidence and most often contradictory of the official news. Such a space of investigation is defined with the same approach as in \cite{bessi2015science, del2016spreading}, with the support of different Facebook groups very active in monitoring the conspiracy narratives. For each page, we downloaded all the posts (and their respective metadata) in a timespan of 5 years (Jan 1, 2010 to Dec 31, 2014). The dataset is composed by $345,054$ posts. To our knowledge, the dataset is the complete set of conspiracy-like information sources active in the US Facebook scenario up to December 31, 2014. 

\subsection{Fundamentals of Extreme Value Theory}
Extreme value theory (EVT) is a branch of statistics dealing with the extreme deviations from the median of probability distributions. In particular, it aims at assessing the probability of events that are more extreme than any previously observed. Extreme value theory is widely used in many fields of science where power laws play a role in modeling \cite{cirillo2015statistical}, such as structural and geological engineering, finance and risk management, earth sciences, traffic prediction, etc.

In this section, we briefly review some fundamental results of extreme value theory. For extended discussion, proofs, and theorems see \cite{gumbel1958statistics, coles2001introduction, embrechts2013modelling}.

\subsubsection{Extreme Value Theory}
Suppose $X_{1},X_{2},\dots$ are independent and identically distributed (iid) random variables with common cumulative distribution function (cdf) $F$. Let $M_{n} = \max\{ X_{1},\dots,X_{n}\}$ denote the maximum of the first $n$ random variables (partial maxima) and let $u(F) = \sup\{ x : F(x) < 1\}$ denote the upper endpoint of $F$. Since
$$ \mathbf{Pr}(M_{n} \leq x) = \mathbf{Pr}(X_{1} \leq x, \dots, X_{n} \leq x) = F^{n}(x),$$

$M_{n}$ converges almost surely to $u(F)$ whether it is finite or infinite. Extreme value theory seeks norming constants $a_{n}>0$, $b_{n} \in \mathbb{R}$, and some nondegenerate distribution function $G$ such that the cdf of the normalized $M_{n}$ converges to $G$, i.e.
$$ \mathbf{Pr}\left( \frac{M_{n} - b_{n}}{a_{n}} \leq x \right) = F^{n}(a_{n}x + b_{n}) \to^{d} G(x).$$

If this holds for suitable choices of $a_{n}$ and $b_{n}$, then we say that $G$ is an extreme value distribution function, and $F$ belongs to the maximum domain of attraction of $G$, i.e. $F \in MDA(G)$. The Extremal Types Theorem characterizes the limit distribution function $G$ as of the type of one of the following three classes:
\begin{itemize}
	\item Gumbel: 
	$$\Lambda(x) = \exp\left(\exp(-x)\right), \quad x \in \mathbb{R}$$
	\item Fr\'{e}chet: 	
	\begin{equation*}
	\Phi_{\alpha}(x)=\begin{cases}
	0, & \text{if $x \leq 0$}\\
	\exp(-x^{-\alpha}), & \text{if $x > 0$}
	\end{cases}
	\end{equation*}
	\item Weibull
	\begin{equation*}
	\Psi_{\alpha}(x)=\begin{cases}
	\exp(-(-x)^{\alpha}), & \text{if $x \leq 0$}\\
	1, & \text{if $x > 0$}
	\end{cases}
	\end{equation*}
	
\end{itemize}
for some $\alpha > 0$.

The three extreme value distributions can be represented using the generalized extreme value (GEV) distribution (family). Let
\begin{equation*}
H_{\xi}(x)=\begin{cases}
\exp\left(-(1 + \xi x)^{-\frac{1}{\xi}}\right), & \text{if $ \xi \neq 0$}\\
\exp(-\exp(-x)), & \text{if $\xi = 0$}
\end{cases}
\end{equation*}
where $1 + \xi x > 0$. Then,
\begin{itemize}
	\item $ \xi = \alpha^{-1} > 0 \longleftrightarrow \Phi_{\alpha}$
	\item $ \xi = -\alpha^{-1} < 0 \longleftrightarrow \Psi_{\alpha}$
	\item $ \xi = 0 \longleftrightarrow \Lambda$
\end{itemize}

From a modeling point of view, the three extreme value distributions are very different, especially for what concerns the behavior of the tails, i.e. the part of the distribution more relevant when dealing with extreme events.

Here, we focus on the Fr\'{e}chet case, $\Phi_{\alpha}$ with $\alpha > 0$. If we consider the tail of $\Phi_{\alpha}(x)$, a Taylor expansion shows that
$$ 1 -\Phi_{\alpha}(x) = 1 - \exp(-x^{-\alpha})  \sim x^{-\alpha},\quad x \to \infty.$$
Hence, $\Phi_{\alpha}(x)$ tends to decrease as a power law. Moreover, every distribution function that belongs to MDA($\Phi_{\alpha}$) has necessarily and infinite right endpoint, i.e. it is defined for $x \in [0,\infty)$. It follows that all the distribution functions belonging to MDA($\Phi_{\alpha}$) are appropriate for modeling phenomena with extremely large maxima.


Consider a random variable $X$ with unknown distribution function $G$ and right endpoint $x_{G} = \sup\{x \in \mathbb{R}: G(x) < 1\}$. The exceedance distribution function of $X$ above a given threshold $t$ is defined as
$$ G_{t}(x) = \mathbf{Pr}(X  \leq x | X > t) = \frac{G(x) - G(t)}{1 - G(t)}, \quad x \geq t.$$

For a large class of distribution functions $G$ and a high threshold $t \to x_{G}$, $G_{t}$ can be approximated by a Generalized Pareto Distribution, i.e.
\begin{equation*}
G_{t} = GPD(x; \xi, \beta, t) = \begin{cases}
1 - \left(  1 + \xi\frac{x - t}{\beta} \right)^{-\frac{1}{\xi}}, & \text{if $ \xi \neq 0$}\\
1 - \exp\left(-\frac{x - t}{\beta}\right), & \text{if $\xi = 0$}
\end{cases}
\end{equation*}
where $x \geq t$ for $\xi \geq 0$, $t \leq x \leq t - \beta/\xi$ for $\xi < 0$, $t \in \mathbb{R}$, $\xi \in \mathbb{R}$, and $\beta > 0$. The shape parameter, $\xi$, governs the fatness of the tails, and thus the existence of the moments. The moment of order $p$ of a Generalized Pareto distributed random variable only exists if and only if $\xi < 1/p$.

\subsubsection{Extreme Value Analysis}
Two approaches exist for practical extreme value analysis. The Block Maxima (BM) approach consists on splitting the observation period into a certain number of non-overlapping periods of equal size -- e.g. weeks, months, years -- and then considering only the maximal value within each period. Such maximal values follow approximately a Generalized Extreme Value (GEV) distribution.

The Peaks Over Threshold (POT) approach relies on considering only the values exceeding a certain high threshold. The probability distribution of those selected observations is approximately a Generalized Pareto Distribution (GPD).

Both approaches have some limitations. The POT approach picks up all relevant high observations, and thus seems to make better use of the available information. Conversely, the BM approach misses some of these high observations and retains some lower observations. However, there may be reason for using the BM method: the only available information may be block maxima (e.g. daily, weekly, monthly, or yearly maxima) and the BM approach may be preferable when the observations are not exactly iid.

However, if the BM approach may be easier to apply when the block periods appear naturally, some problems arise when this does not happen. In such a case, the choice of the block size for the BM approach may be as difficult as the choice of the threshold for the POT approach. 

\begin{figure*}[ht]
	\centering
	\includegraphics[width = \textwidth]{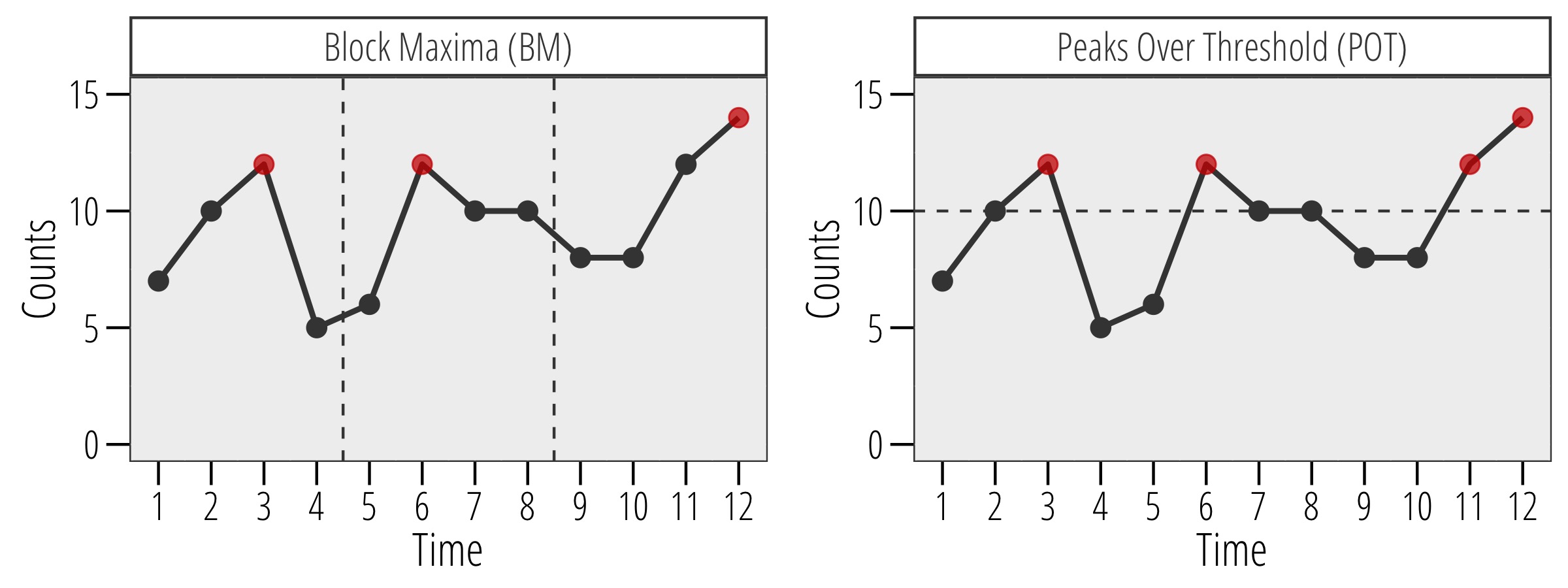}
	
	\caption{\textbf{Block Maxima (BM) vs. Peaks Over Threshold (POT).}}
	\label{fig:0}
\end{figure*}

Figure \ref{fig:0} provides a graphical representation of the two approaches.

\section{Results and Discussion}
In this paper, we aim at investigating the statistical properties of viral misinformation on Facebook by means of extreme value theory. More specifically, the object of the analysis is the number of times that posts supporting --- in this study, we are assuming that each share of conspiracy posts represents the will to support a given conspiracy narrative --- conspiracy theories have been shared, which can be considered as a random variable $X$ following a generic distribution function $F$ with support $[0,\infty)$. Such an infinite right endpoint is justified by the fact that users can share a post how many times they desire.

\subsection{Exploratory Data Analysis}
Since for each post we know the time of creation, we have a temporally ordered collection of observations. Such a time series is irregularly spaced, in the sense that it is characterized by varying interarrival times between observations. A common approach to analyze irregularly spaced time series consists in transforming the data into equally spaced observations using interpolation methods, and then apply standard methods for equally spaced data. However, such a transformation can introduce a number of significant and hard to quantify bias, especially when the interarrival times between observations are highly irregular. Since in our case the spacing of observations varies from seconds to days, we avoid to transform data. Moreover, despite observations are temporally ordered, it is difficult to assume some kind of time dependence between the number of shares received by posts --- i.e. the number of shares received by  a post does not affect the number of shares received by following posts. Rather, if we can conceive that some external events (e.g. breaking news, top stories, scandals, etc.) can cause an unusual number of posts in a restricted temporal window (clustering), we can safely assume that the number of shares received by each of those posts is independent and identically distributed (iid).

\begin{figure*}[ht]
	\centering
	\includegraphics[width =\textwidth]{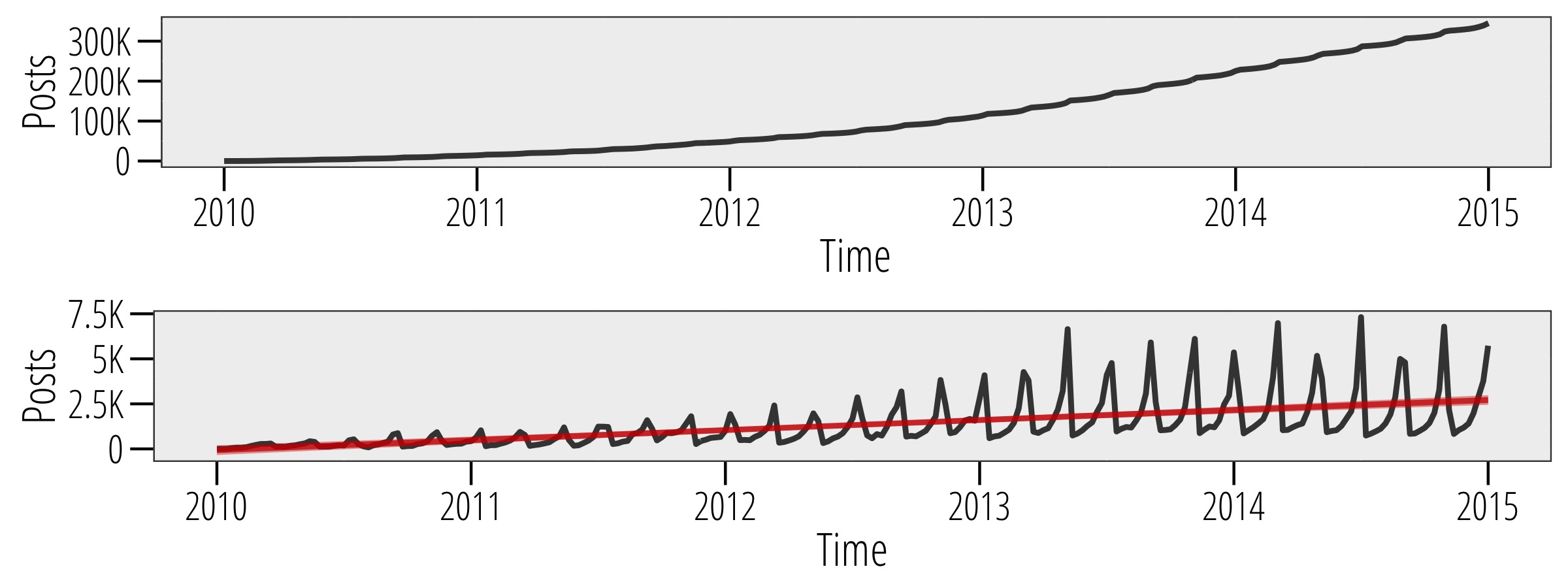}
	
	\caption{\textbf{Weekly posts.} Cumulative number of weekly post (\emph{top panel}) and number of weekly posts (\emph{bottom panel}). The solid red line indicates the fitted linear trend.}
	\label{fig:1}
\end{figure*}

Figure \ref{fig:1} shows the cumulative number of weekly post (\emph{top panel}) and the number of weekly posts (\emph{bottom panel}). Despite it looks like there is a clear growth trend --- which is likely due by the increase of users on Facebook occurring from 2010 to 2014 --- and some form of seasonality, we are not able to identify any meaningful seasonality pattern. Indeed, we can assume that the activity of this kind of pages is primarily driven by external events, such as breaking news, top stories, scandals, etc.

However, an increase in the number of posts published by pages in a given temporal window may reflect an increase in the users' excitement and activity. We account for such a possible characteristic of the phenomenon under investigation by rescaling raw data by a factor defined as

$$ R_{i} = \frac{w_{i}}{\max(w)}, \quad i \in \{1,\dots,261 \}$$

where $w_{i}$ represents the number of posts published by pages in week $i$. Such a rescaling factor inflates the number of shares of posts published in weeks characterized by an overall low activity. Different rescaling strategies have been considered --- e.g. rescaling by the mean or the median ---, and similar results have been obtained.

\begin{figure*}[ht]
	\centering
	\includegraphics[width =\textwidth]{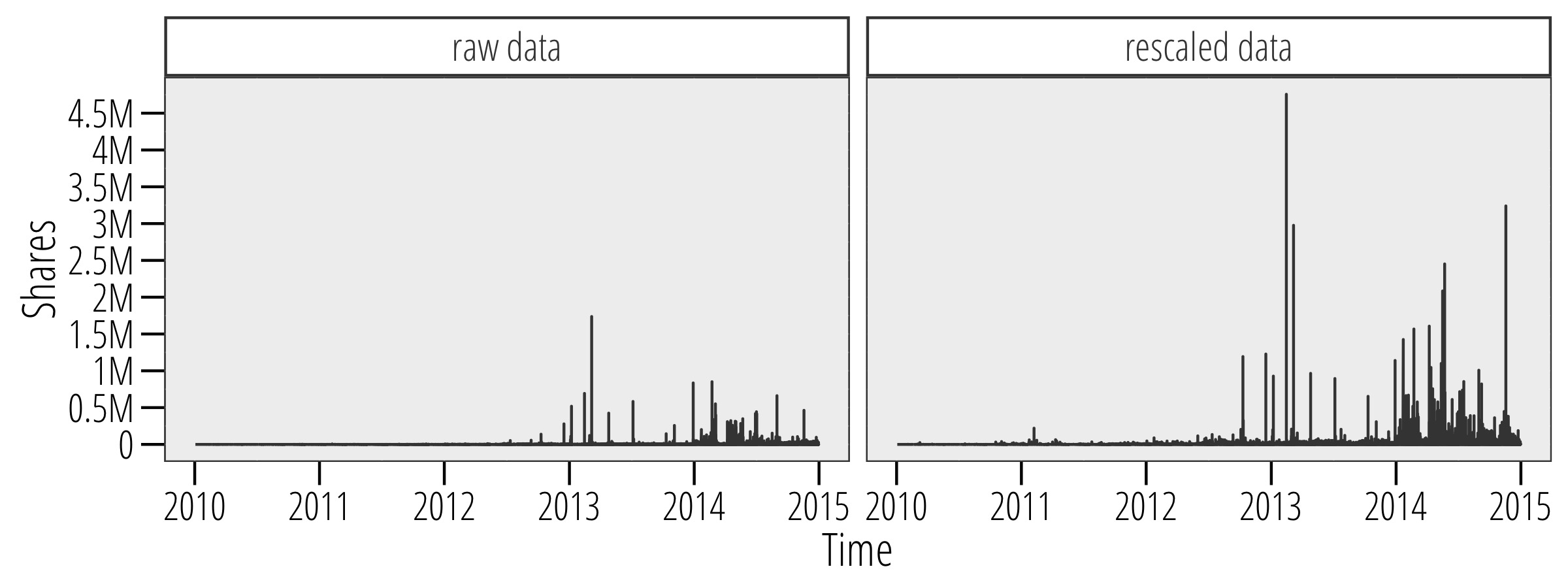}
	
	\caption{\textbf{Shares time series.} Time series of the original number of shares (\emph{raw data}) and the rescaled one (\emph{rescaled data}).}
	\label{fig:2}
\end{figure*}

\begin{figure*}[ht]
	\centering
	\includegraphics[width = \textwidth]{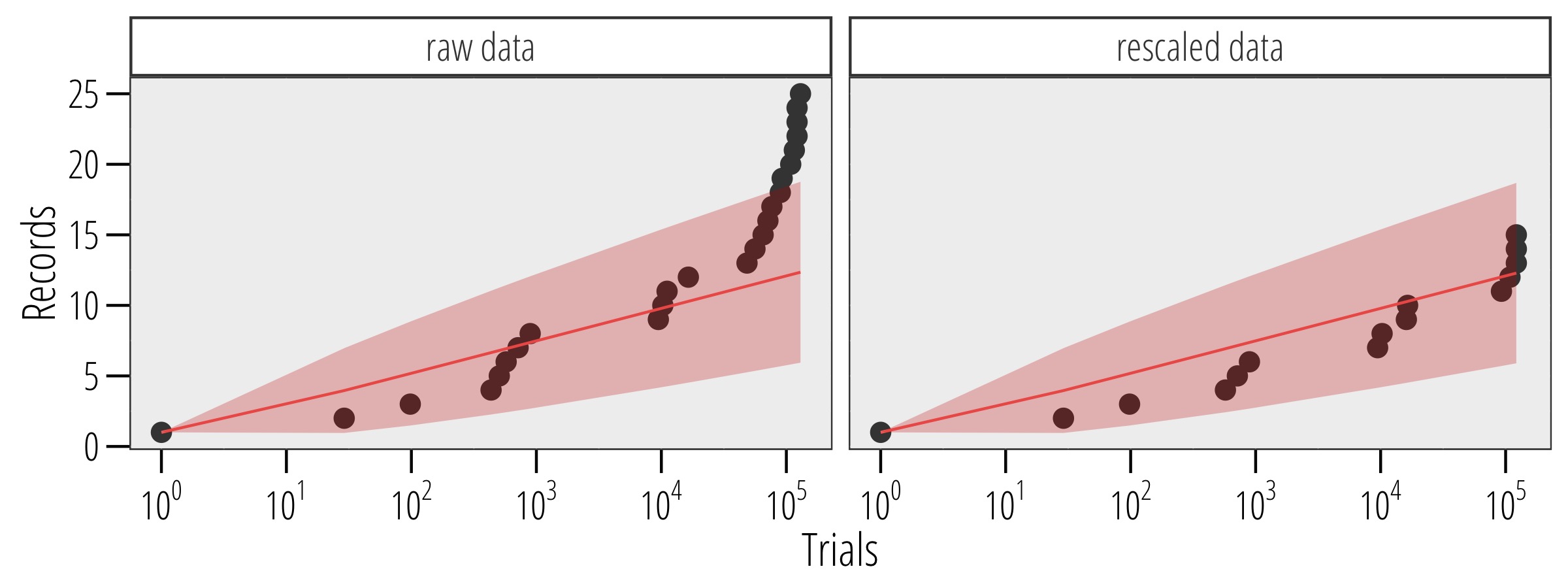}
	\caption{\textbf{Records plots.} The iid assumption for raw data is violated, but still valid for rescaled data, where records are distributed around their expected value and within the $95\%$ confidence intervals.}
	\label{fig:3}
\end{figure*}

Figure \ref{fig:2} shows the time series of the original number of shares (\emph{raw data}) and the rescaled one (\emph{rescaled data}). The rescaling procedure should have removed or at least reduced possible clustering phenomena that would have led to a violation of the iid assumption. We check the iid assumption by means of the records plot, a simple and intuitive exploratory tool widely used in extreme value analysis which exploits the fact that successive records for iid data should become more and more rare as time goes by. Since a record $x_{n}$ for the random variable $X$ occurs if $ x_{n} > \max\{ x_{1}, \dots, x_{n-1} \}$, it is intuitive that if data are iid it becomes more difficult to exceed all past observations, and thus the number of records should follow a logarithmic pattern \cite{gumbel1958statistics}. Records plots in Figure \ref{fig:3} show that the iid assumption for raw data is violated, but still valid for rescaled data, where records are distributed around their expected value and within the $95\%$ confidence intervals.

\begin{figure*}[ht]
	\centering
	\includegraphics[width = \textwidth]{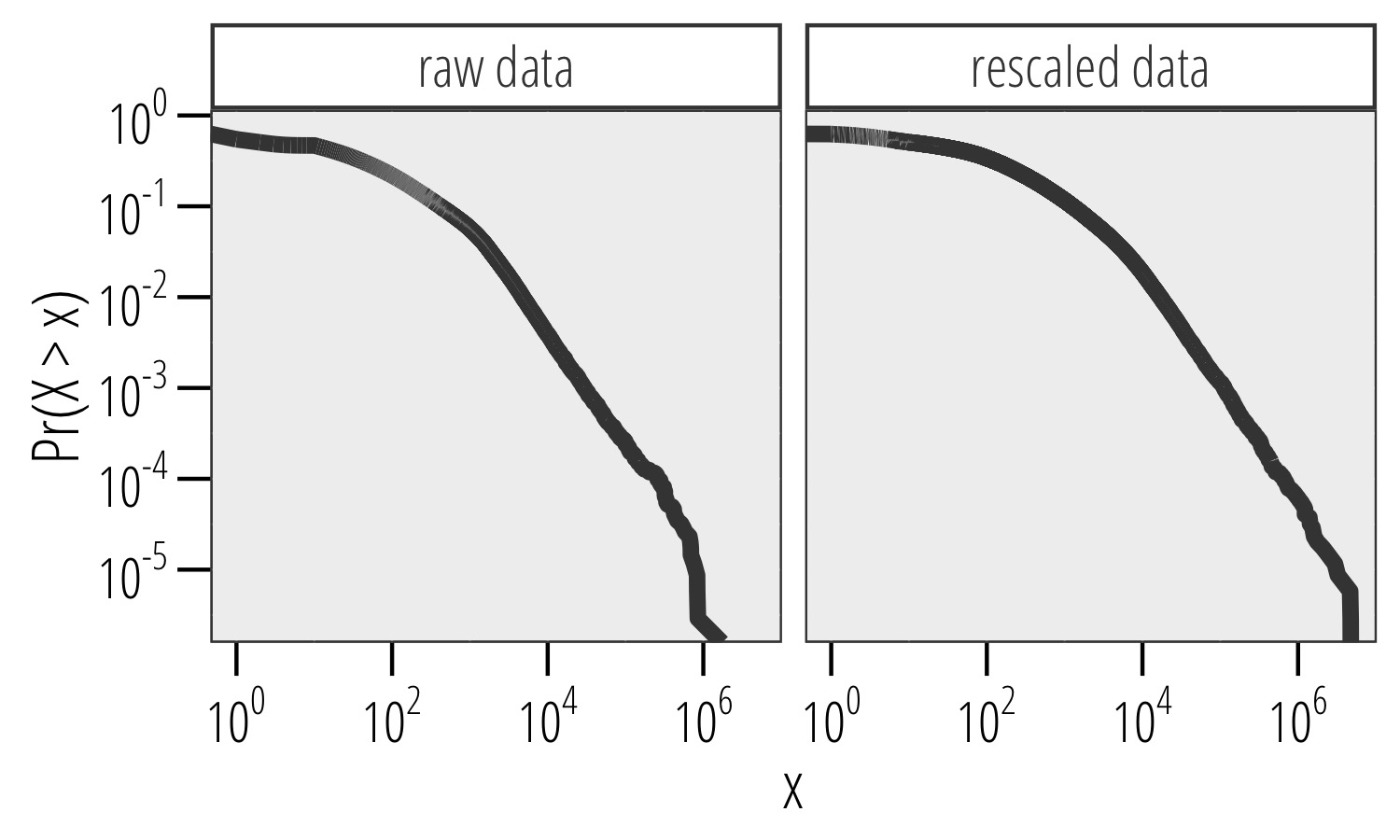}
	\caption{\textbf{Empirical Complementary Cumulative Distribution Function.} The rescaling procedure slightly exacerbates the power law behavior of tail without influencing the body of the distribution.}
	\label{fig:4}
\end{figure*}

Figure \ref{fig:4} shows the empirical complementary cumulative distribution functions of raw data and rescaled data. The double log scale of the figures highlights the fatness of the right tail in both the two empirical distributions. Beyond removing any form of dependence in the raw data, we observe that the rescaling procedure slightly exacerbates the power law behavior of the tail without influencing the body of the distribution. Thus, rescaled data will be used for successive analysis.

\subsection{Statistical Properties of Viral Misinformation}
We use the distribution function of rescaled data to characterize the statistical properties of viral misinformation by means of EVT tools. First, we analyze the limit behavior of the Maximum/Sum ratio 
$$ R_{n}(p) = \frac{M_{n}(p)}{S_{n}(p)}, \qquad n\geq 1, p > 0,$$

where $S_{n}(p) = \sum^{n}_{i = 1}(X_{i}^{p})$ and $M_{n}(p) = \max(X_{i}^{p})$. The moment of order $p$ of the distribution exists, i.e. $E[X^{p}] < \infty$, if and only if $R_{n}(p)$ converges to zero for $n \to \infty$. Conversely, an erratic limit behavior of $R_{n}(p)$ indicates the infiniteness of the $p$-th moment of the distribution, i.e. $E[X^{p}] = \infty$.

\begin{figure*}[ht]
	\centering
	\includegraphics[width = \textwidth]{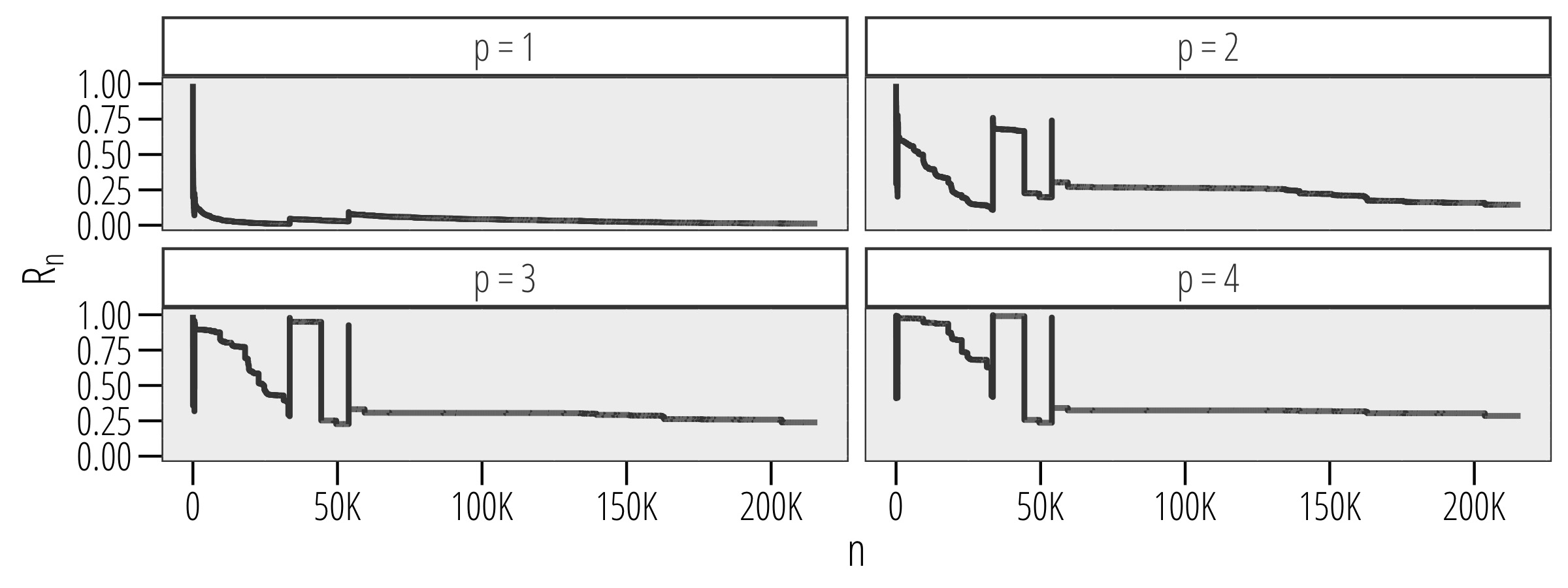}
	
	\caption{\textbf{Maximum/Sum ratio plot.} Only the first moment of the distribution exists, i.e. $E[X] < \infty$, whereas moments of order greater than $p = 2$ are infinite, i.e. $E[X^{p}] = \infty$ for $p\geq 2$.}
	\label{fig:5}
\end{figure*}

Figure \ref{fig:5} shows that only the first moment of the distribution exists, i.e. $E[X] < \infty$, whereas moments of order greater than $p = 2$ are infinite, i.e. $E[X^{p}] = \infty$ for $p\geq 2$. Identical results hold for raw data. Our distribution function belongs to the maximum domain of attraction of Fr\'{e}chet, i.e. $F \in MDA(\Phi_{\alpha})$. The existence of the first moment of the distribution function allows us to compute a \emph{reasonable} (in a sample one can compute basically anything, even meaningless quantities) estimate of the conditional tail mean above a given threshold. Indeed, by the law of total expectation

$$E[X| X > t] = \frac{E[X] - \mathbf{Pr}(X \leq t) E[X| X \leq t]}{\mathbf{Pr}(X > t)},$$

where $E[X| X \leq t]$ is finite since bounded from above by $t$, and the finiteness of $E[X]$ implies that the conditional tail mean, $E[X| X > t]$, is finite.

Such a measure is known in finance as the expected shortfall of a loss distribution, and it let us answer to question such as \emph{``What is the expected number of shares for a post once it has exceeded the $250K$ shares threshold?"}. Indeed,
$$ E[X | X > t] = \frac{\sum_{i = 1}^{N}x_{i}\mathbb{I}(x_{i} > t)}{\sum_{i = 1}^{N}\mathbb{I}(x_{i} > t)} = \frac{\sum_{i = 1}^{N}x_{i}\mathbb{I}(x_{i} > 250K)}{\sum_{i = 1}^{N}\mathbb{I}(x_{i} > 250K)} \approx 467K.$$ 

Since we showed that the moments of order greater than $1$ do not exist, one should prefer the mean absolute deviation over the variance as a measure of dispersion around the conditional tail mean. 

Recall that the moment of order $p$ of a Generalized Pareto distributed random variable only exists if and only if $\xi < 1/p$, and thus the shape parameter we are going to estimate can not be smaller than $1/2$. Such an observation has the main implication that we can safely use the maximum likelihood (ML) approach to estimate $\xi$, since the ML estimates are consistent only when $\xi > -1/2$. 

Since our time series is highly irregularly spaced, with interarrival times ranging between seconds and days, we prefer the Peaks Over Threshold (POT) approach over the Block Maxima (BM) approach to estimate the shape parameter $\xi$. Indeed, when the block periods used in the BM approach does not appear naturally, the choice of a threshold for the POT approach may be easier.

\begin{figure*}[ht]
	\centering
	\includegraphics[width = \textwidth]{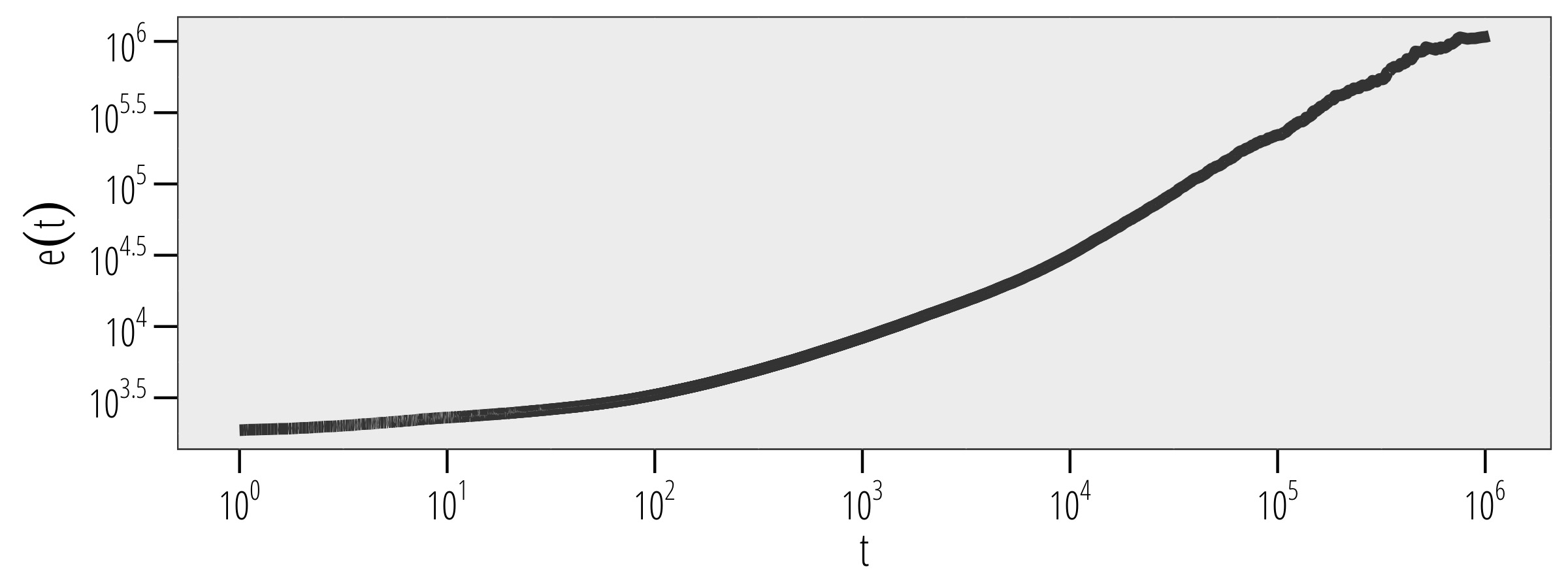}
	
	\caption{\textbf{Empirical Mean Excess Function plot.}}
	\label{fig:6}
\end{figure*}

Before fitting the distribution function to a Generalized Pareto Distribution, we have to identify a feasible threshold. To accomplish such a task, we rely on the Mean Excess Function (MEF). The empirical MEF of a sample of observations $x_{1},\dots,x_{n}$ is defined as
$$ e_{n}(t) = \frac{\sum_{i = 1}^{n}(x_{i} - t)}{\sum_{i=1}^{n}\mathbb{I}(x_{i} > t)}, $$

that is the ratio between the sum and the number of the exceedances over the threshold $t$.
Figure \ref{fig:6} shows the MEF plot for the rescaled data. We observe that the empirical MEF begins to linearly increase in the threshold at $t \approx 10^{4}$. Such a behavior characterizes power law distribution functions \cite{cirillo2013your}, and thus we choose $t = 10^{4}$ as threshold.

\begin{figure*}[ht]
	\centering
	\includegraphics[width = \textwidth]{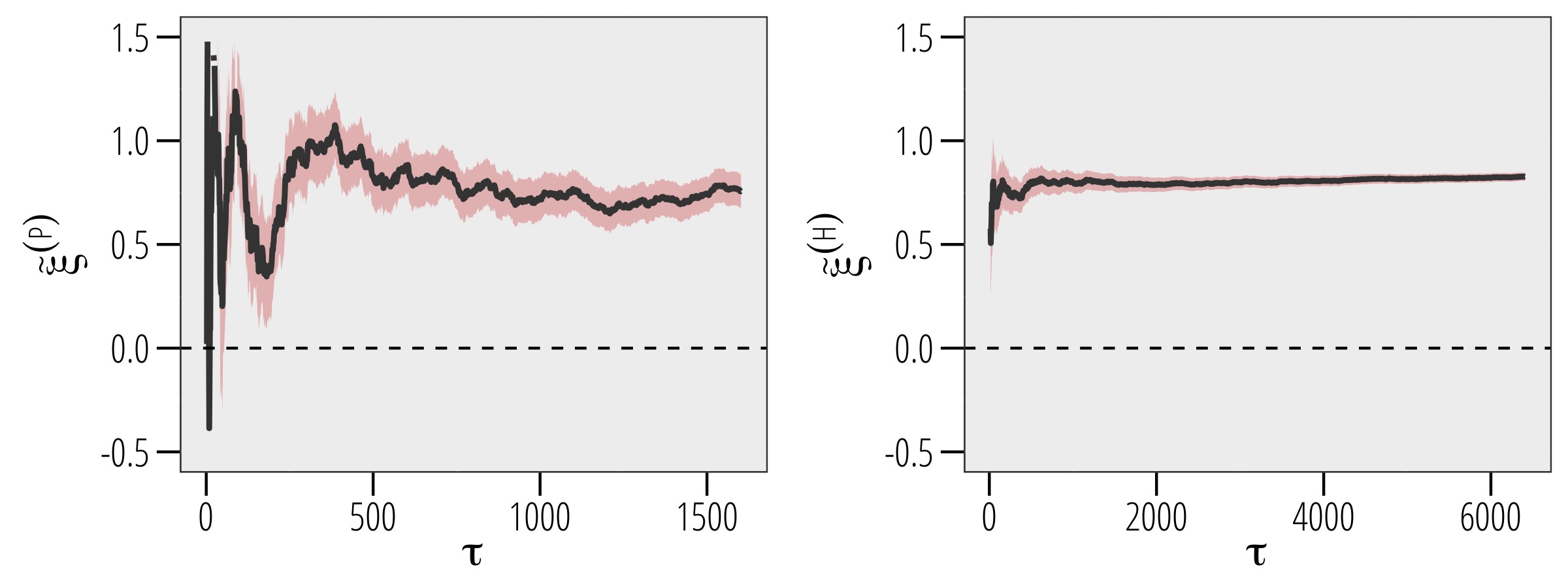}
	
	\caption{\textbf{Pickands and Hill nonparametric estimators for the shape parameter $\xi$.}}
	\label{fig:7}
\end{figure*}

Given the heavy-tailed behavior of the rescaled data distribution function, the shape parameter $\xi$ is likely to be positive. The left panel of Figure \ref{fig:7} shows the Pickands plot, based on the nonparametric Pickands estimator for $\xi$, defined as 
$$ \tilde{\xi}^{(P)}_{\tau,n} = \frac{1}{\log_{2}} \log\frac{X_{\tau,n} - X_{2\tau,n}}{X_{2\tau,n} - X_{4\tau,n}},\qquad \tau = 1,\dots, \lfloor n/4 \rfloor$$

where $X_{\tau,n}$ is the $\tau$-th upper order statistics out of a sample of $n$ observations. The Pickands plot shows a more or less stable behavior of the Pickands estimates for different values of $\tau$, suggesting that the true value of $\xi$ lies in the interval $(0.5,1)$.

The right panel of Figure \ref{fig:7} shows the Hill plot, based on the nonparametric Hill estimator for $\xi$, defined as
$$ \tilde{\xi}^{(H)}_{\tau,n} = \frac{1}{\tau} \sum_{j = 1}^{\tau} \ln(X_{j,n}) - \ln(X_{\tau,n}),$$

where $X_{j,n}$ is the $j$-th upper order statistics out of a sample of $n$ observations. The Hill plot outperforms the Pickands plot in stability, suggesting a true value of $\xi$ around $0.75$. 

The main implication is that, as already anticipated by the analysis of the Maximum/Sum ratio plot, the true value of $\xi$ is greater than $-1/2$, and thus we can obtain consistent estimates via a maximum likelihood (ML) approach. Table \ref{tab:1} shows ML estimates of $\xi$ and $\beta$ for different thresholds. We observe a stable value of $\tilde{\xi}^{ML}$ for increasing values of the threshold. We obtain similar results for raw data (i.e. $\tilde{\xi}^{ML} = 0.769 (0.0198)$ with $t = 2.5K$).

\begin{table}[ht]
	\centering
	\caption{\textbf{Maximum Likelihood estimates, standard errors, and number of exceedances for different thresholds.}}
	\label{tab:1}
	\begin{tabular}{|c | c | c | c| }
		threshold & $\tilde{\xi}^{(ML)}$ & $\tilde{\beta}^{(ML)}$ & $\#$ exceedances \\
		\hline
		$10K$ & $0.770$ & $8,750$ & $6,408$ \\
		& ($0.0220$) &  ($205$)& \\
		
		$25K$ & $0.800$  & $19,500$ & $2,153$\\
		& ($0.0391$) &  ($ 805$)& \\
		
		$50K$ & $0.737$  & $ 43,170$ & $884$\\
		& ($0.059$) &  ($2,730$)& \\
		
		$100K$ & $0.746$  & $74,380$ & $399$\\
		& ($0.0869$) &  ($6,923$)& \\
		
		$150K$ & $0.726$  & $120,460$ & $223$\\
		& ($0.120$) &  ($15,500$) & \\
		\hline
	\end{tabular}
\end{table}

\subsection{Frequency of Viral Misinformation}
The Peaks Over Threshold (POT) method has two main implications: the exceedances over a high threshold follow a Generalized Pareto Distribution, and the number of excesses over time follows a homogeneous Poisson process. In a homogeneous Poisson process the number of events, $N(\theta)$, in a finite interval of time of length $\theta$ follows the Poisson distribution, i.e.
$$ \mathbf{Pr}(N(\theta) = n) = \frac{(\lambda\theta)^{n}}{n!}\exp(-\lambda\theta).$$

Moreover, the interarrival times between events are independent and follow the exponential distribution, i.e.
$$ \mathbf{Pr}(\text{interarrival time} > \theta) = \exp(-\lambda\theta). $$

\begin{figure*}[ht]
	\centering
	\includegraphics[width = \textwidth]{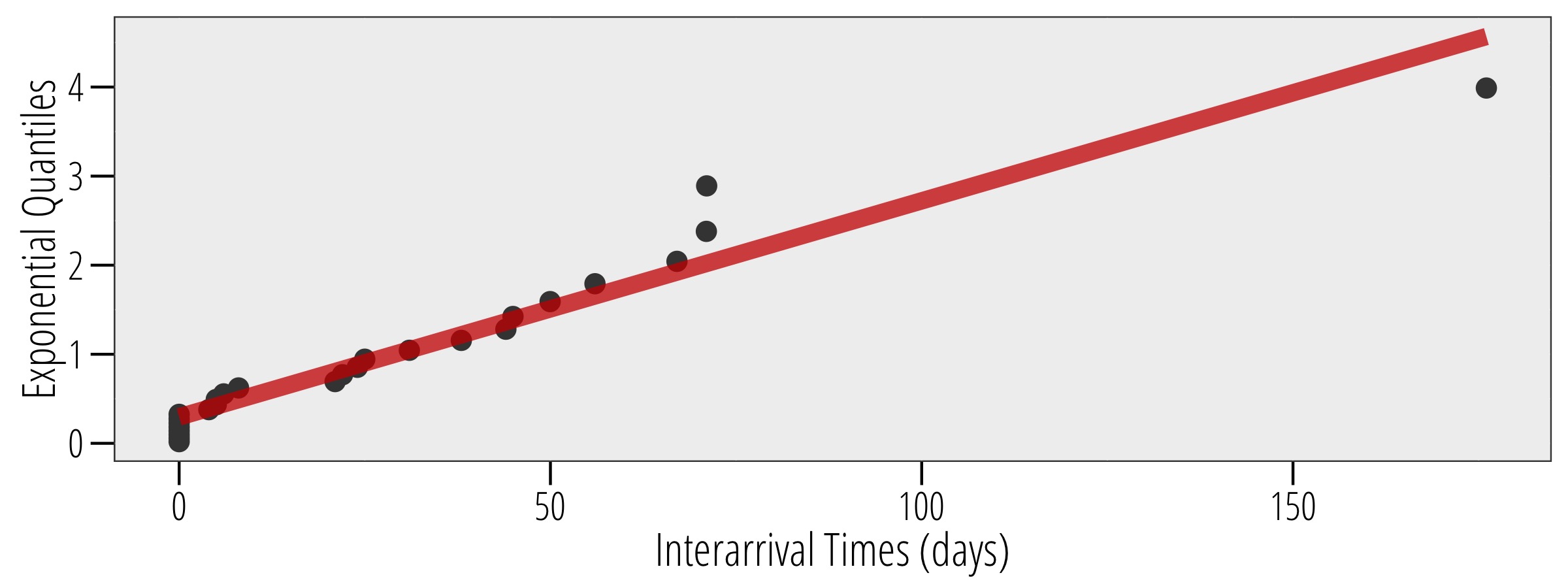}
	
	\caption{\textbf{Exponential Quantiles vs. Interarrival Times.}}
	\label{fig:8}
\end{figure*}

\begin{figure*}[ht]
	\centering
	\includegraphics[width = \textwidth]{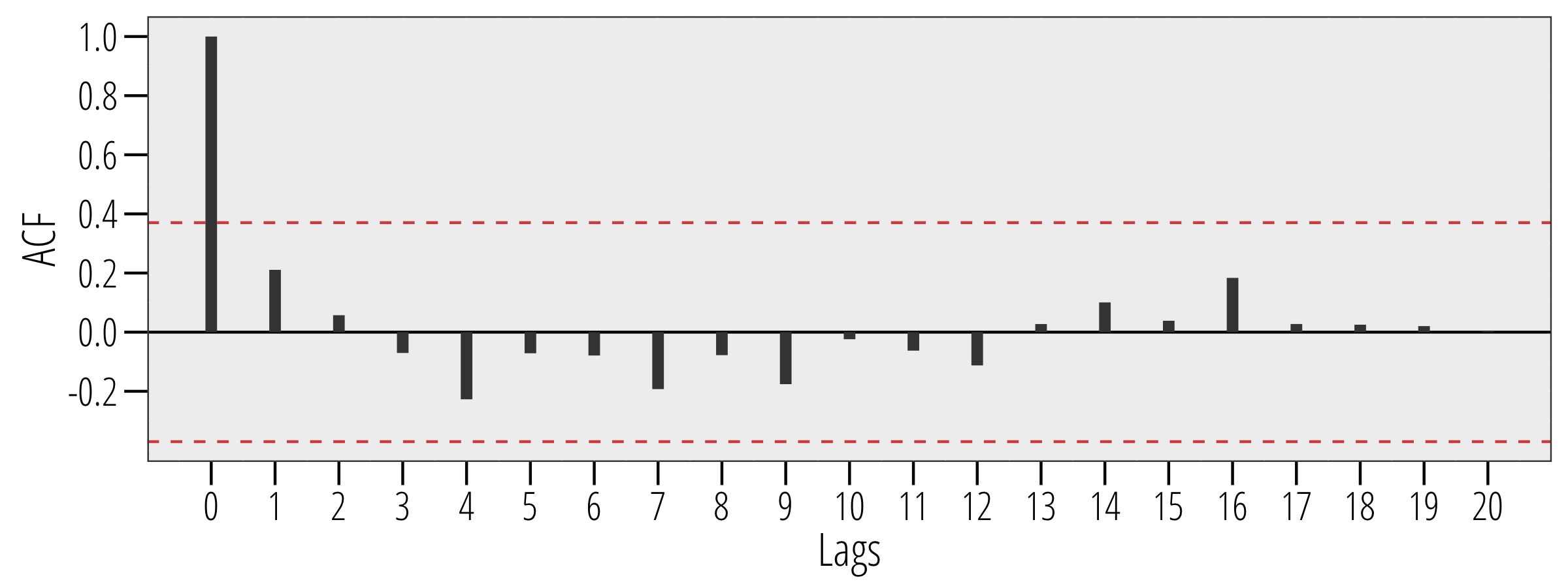}
	
	\caption{\textbf{Autocorrelogram for Interarrival Times.} We find no correlation as a function of the time lag between them.}
	\label{fig:9}
\end{figure*}

Figure \ref{fig:8} shows that the interarrival times of posts shared more than $750K$ times (rescaled data) follow approximately an exponential distribution. Moreover, the autocorrelogram function (ACF) plot ---  i.e. a plot showing the similarity between observations as a function of the time lag between them \cite{box2015time} --- in Figure \ref{fig:9} shows that the interarrival times between those posts are independent, supporting the i.i.d. hypothesis suggested by the records plot in Figure \ref{fig:3}. Similar results approximately hold for raw data when considering posts shared more than $250K$ (the bootstrap test of fit for the Generalized Pareto Distribution \cite{villasenor2009bootstrap} gives a p-value equal to $0.2$). We conclude that the number of extremely viral posts over time follows a homogeneous Poisson process.

Such a conclusion allows us to exploit some useful properties of the homogeneous Poisson processes to quantify the frequency of rare viral contents on online social media. Indeed, the expected value of the number of events, $N(\theta)$, in a finite interval of time of length $\theta$ is defined as
$$ E[N(\theta)] = \lambda\theta,$$
where $\lambda > 0$ is known as the rate parameter of the Poisson process. The reciprocal of such a parameter, i.e. $1/\lambda$, is known as the survival parameter of the exponential distribution followed by the interarrival times between the $N(\theta)$ events. Given a sample $z_{1}, \dots, z_{n}$ of interarrival times, the survival parameter is estimated through the sample mean
$$ \frac{1}{\lambda} = \frac{\sum_{i = 1}^{n}z_{i}}{n}.$$

Essentially, we can estimate the survival parameter, $1/\lambda$, of the exponential distribution describing the interarrival times between rare events exceeding a certain threshold, and then use the rate parameter, $\lambda$, of the Poisson process to estimate the expected number of events exceeding that threshold in a finite time of length $\theta$.

For instance, the survival parameter of the interarrival times distribution of posts exceeding $250K$ shares (raw data) is $1/\lambda = 18.5$. It follows that $\lambda = 1 / 18.5 = 0.0541$. Basically, if by means of the survival parameter we can answer to questions such as \emph{``What is the mean waiting time between posts exceeding $250K$ shares?"}, through the rate parameter we can answer to questions such as \emph{``What is the expected number of posts exceeding $250K$ shares in the future $365$ days?"}. Indeed,
$$ E[N(\theta)] = \lambda\theta = 0.0541 \times 365 = 19.8 \approx 20.$$

A convenient way to assess the uncertainty around the rate parameter, $\lambda$, consists in using a standard Bayesian probability updating method. Indeed, the conjugate prior distribution for a Poisson distribution is the Gamma distribution, and we can express the prior distribution of $\lambda$ as
$$ \mathbf{Pr}(\lambda) = \text{Gamma}(\alpha,\beta).$$

\begin{figure*}[ht]
	\centering
	\includegraphics[width = \textwidth]{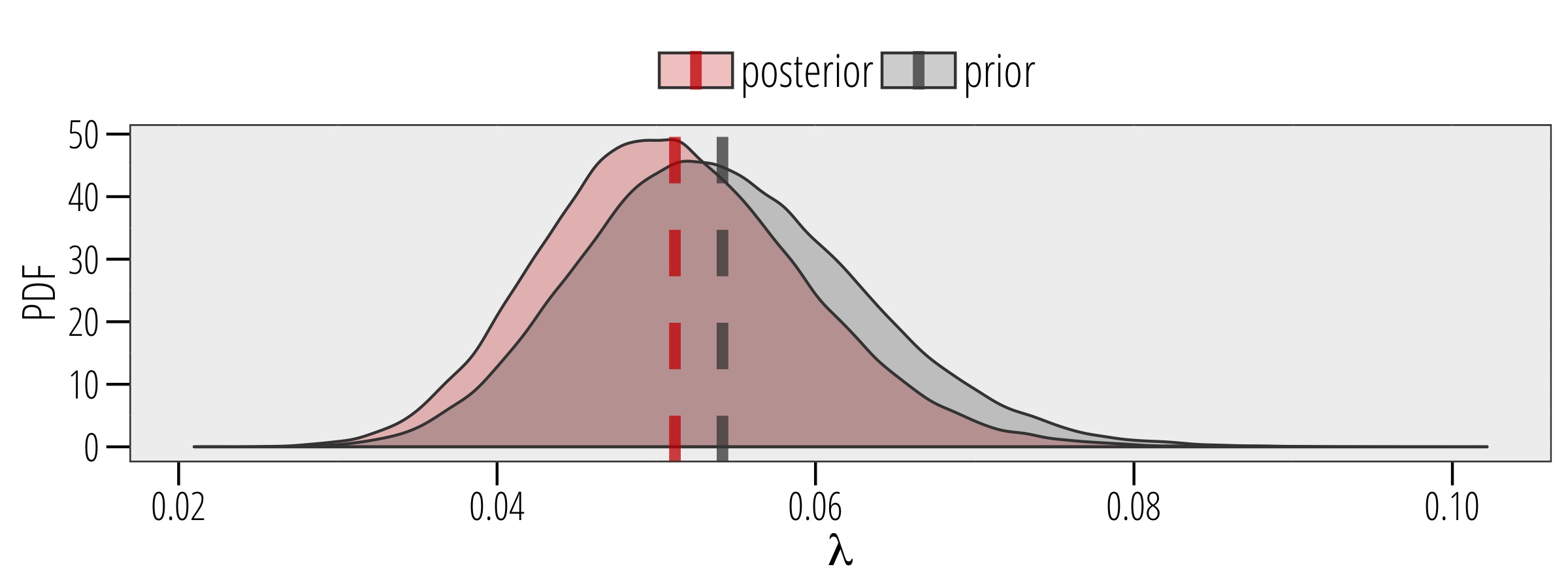}
	
	\caption{\textbf{Prior and Posterior Distributions of the rate parameter.} The grey and red dashed lines indicate, respectively, the mean of the prior and the mean of the posterior distribution of the rate parameter. }
	\label{fig:10}
\end{figure*}

Since the expected value (mean) of a Gamma distribution is defined as $\alpha/\beta$, we may want to choose the hyperparameters, $\alpha$ and $\beta$, of the prior distribution $ \mathbf{Pr}(\lambda)$ so that
$$ \frac{1}{\lambda} = \frac{\beta}{\alpha} =  \frac{\sum_{i = 1}^{n}z_{i}}{n},$$

where $z_{1},\dots,z_{n}$ are the observed interarrival times. Then, the posterior distribution of the rate parameter is defined as
$$ \mathbf{Pr}(\lambda | \mathbf{z}) = \text{Gamma}(\alpha + k, \beta + \sum^{k}_{i = 1}z_{i}),$$

where $z_{1},\dots,z_{k}$ represent $k$ new observed interarrival times. For instance, we may define the prior distribution of the rate parameter of the interarrival times distribution of posts exceeding $250K$ shares (raw data) as
$$ \mathbf{Pr}(\lambda) = \text{Gamma}(\alpha,\beta) = \text{Gamma}(n,\sum_{i = 1}^{n}z_{i}) = \text{Gamma}(38,702), $$

with mean equal to $\alpha / \beta = 38 / 702 = 0.0541$, and variance equal to $\alpha / \beta^{2} = 38 / 702^{2} = 7.71 \times 10^{-5}$. Then, if after $60$ days we observe a post exceeding the $250K$ shares threshold, the posterior probability distribution of the rate parameter is
$$ \mathbf{Pr}(\lambda | \mathbf{z}) = \text{Gamma}(\alpha + k, \beta + \sum^{k}_{i = 1}z_{i}) = \text{Gamma}(38 + 1, 702 + 60),$$

with mean equal to $(\alpha + k) / (\beta +\sum^{k}_{i = 1}z_{i})  = (38 + 1) / (702 + 60) = 0.0512$, and variance equal to $38 + 1 / (702 + 60)^{2} = 6.72 \times 10^{-5}$. Figure \ref{fig:10} shows both the prior and the posterior distributions of the rate parameter in the aforementioned example.

\begin{figure*}[ht]
	\centering
	\includegraphics[width = \textwidth]{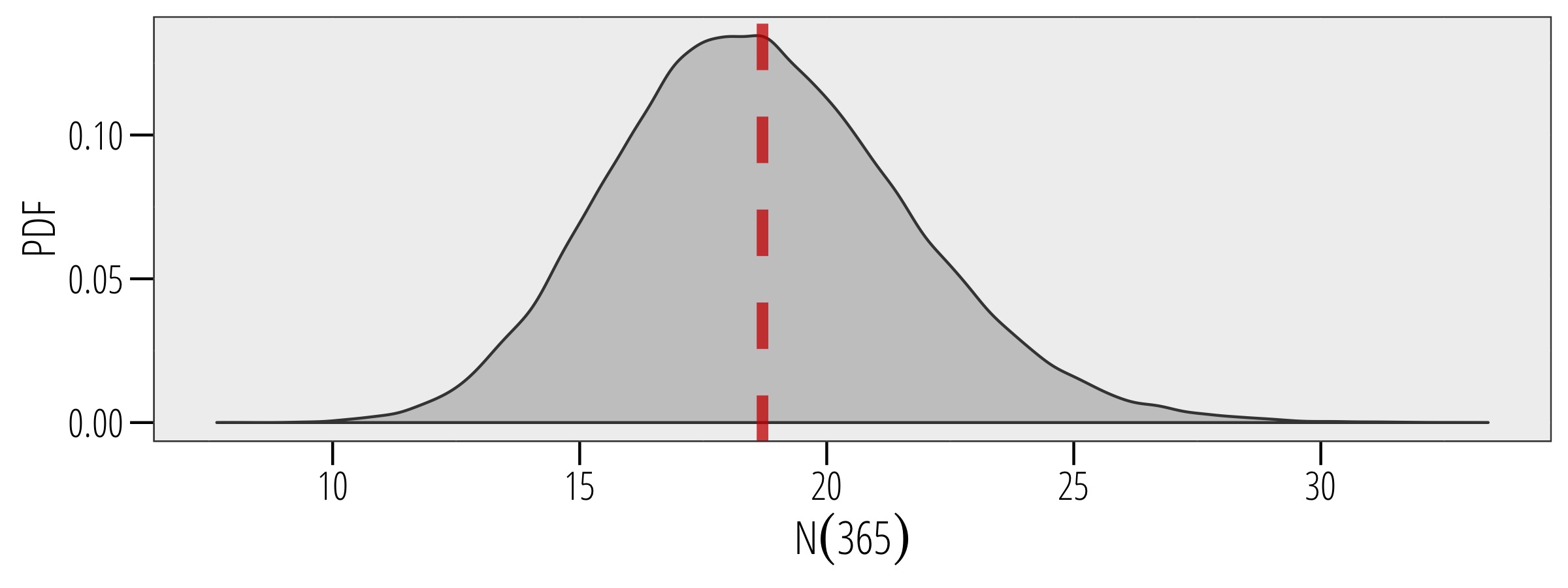}
	
	\caption{\textbf{Posterior predictive probability distribution.} Posterior predictive probability distribution of the number of posts exceeding the $250K$ shares threshold in the next $365$ days. The red dashed line indicates the mean, i.e. $18.7$.}
	\label{fig:11}
\end{figure*}

After such an update, the expected value of the number of posts exceeding the $250K$ threshold in the next $365$ days is defined as
$$ E[N(\theta)] = E[\lambda|\mathbf{z}]\theta = 0.0512 \times 365 = 18.7 \approx 19,$$

and the mean waiting time between posts exceeding the $250K$ shares is $1 / 0.0512 = 19.5$ days. 
Moreover, the full probability assessment of the uncertainty around the rate parameter, $\lambda$, allows us to express the predictive posterior probability distribution of the number of posts exceeding a certain threshold over a finite interval of time

$$ \mathbf{Pr}(N(\theta)|\lambda) = \mathbf{Pr}(\lambda|\mathbf{z}) \theta.$$

Figure \ref{fig:11} shows the posterior predictive probability distribution function of the number of posts exceeding the $250K$ shares threshold (raw data) in the finite interval time of length $365$ days.

\subsection{Concluding Remarks}
In this paper, we study the statistical properties of viral misinformation in online social media. In particular, we focus our attention on Facebook posts spreading false news, hoaxes and unsubstantiated claims. By means of an Extreme Value Theory approach, we show that the number of extremely viral posts over time follows a homogeneous Poisson process, and that the interarrival times between such posts are independent and identically distributed, following an exponential distribution. Moreover, we characterize the uncertainty around the rate parameter of the Poisson process through Bayesian methods. Finally, we are able to derive the predictive posterior probability distribution of the number of posts exceeding a certain threshold of shares over a finite interval of time.

The relevance of our results is not necessarily limited to the field of computational social science coping with misinformation. Despite the prediction of extremely viral posts --- and, more generally, rare events --- remains an hard task, we believe that both our findings and the methodology introduced in this paper may be of interest to the broader field of computational social science dealing with forecasting and tracking of viral contents and events --- e.g. cyber-security attacks, terrorist attacks, etc.

\section*{Acknowledgements}
Special thanks to Geoff Hall and Skepti Forum for providing fundamental support in defining the atlas of Facebook pages disseminating conspiracy theories and myth narratives.

\section*{References}

\bibliography{evt_biblio_revised}

\begin{thebibliography}{10}
\bibitem{coscia2014average}
M.~Coscia, Average is boring: How similarity kills a meme's success, Scientific
  reports 4.
\end{thebibliography}

\end{document}